\documentstyle[11pt,aasms4]{article}

\received{}
\accepted{}
\journalid{}{}
\articleid{}{}

\slugcomment{}

\lefthead{Stocke \& Rector}
\righthead{An Excess of Mg II Absorbers in BL Lac Objects}

\begin{document}

\title{An Excess of Mg II Absorbers in BL Lac Objects}

\author{John T. Stocke\altaffilmark{1,2} and Travis A. Rector\altaffilmark{1,2}}
\affil{Center for Astrophysics and Space Astronomy}
\affil{University of Colorado, Boulder, CO 80309-0389}

\altaffiltext{1}{Visiting Astronomer, Kitt Peak National Observatory. 
KPNO is operated by AURA, Inc.\ under contract to the National Science
Foundation.} 
\altaffiltext{2}{Visiting Astronomer, Multiple Mirror Telescope
Observatory (MMTO), jointly operated by the University of Arizona and the
Smithsonian Institution.} 

\begin{abstract} 
Two new Mg II absorbers are presented ($z=1.340$ in S5 0454+844 and $z=1.117$
in PKS 2029+121), bringing the total number of Mg II systems in the 1 Jy
radio-selected BL Lac sample to 10. Five of the ten absorption systems
are at $W_{\lambda} > 1$\AA; this is a factor of four to five greater than the number expected based upon quasar sightlines, and is 2.5 to 3$\sigma$ greater than the expectation value.  Interpretations of this possible excess
include either that some of the Mg II absorbers might be intrinsic to
the BL Lac or that there is a correlation between the presence of
absorbing gas in the foreground and the nearly featureless spectra of
these BL Lac Objects compared to quasars. Such a correlation can be
created by gravitational microlensing as suggested by Ostriker \&
Vietri.  The similarity between the optical spectra of BL Lacs with 
Mg II absorption and the spectrum of the $\gamma$-ray burst source 
GRB 970508 suggests that models of $\gamma$-ray bursts as microlensed 
AGN should be investigated.
\end{abstract}


\section{Introduction}
BL Lac Objects are thought to be low-luminosity radio galaxies
(Fanaroff-Riley 1974 class 1 sources) whose jet synchrotron and
inverse-Compton emissions are Doppler-boosted (Blandford \& Rees 1978;
Urry \& Padovani 1995). The detailed properties of many (but not all) BL
Lacs are consistent with this scenario, including: (1) their presence
in luminous early-type host galaxies (Abraham, Crawford \& McHardy 1992;
Wurtz, Stocke \& Yee 1996) in poor clusters of galaxies (Wurtz et al.
1997), (2) low-luminosity extended radio structures that often (but not
always) resemble FR 1s (Antonuuci \& Ulvestad 1985; Murphy, Browne \&
Perley 1993; Perlman \& Stocke 1993) and (3) overall spectral energy
distributions (Sambruna, Maraschi \& Urry 1996) and luminosity functions
(Urry \& Padovani 1995) consistent with Doppler-boosted FR 1 energy
distributions and luminosity functions. The multi-wavelength properties
of X-ray selected BL Lacs (XBLs; Stocke et al. 1991) are especially
well-described by the beamed FR 1 model. However, many of the radio-selected 
BL Lacs (RBLs) have somewhat different properties, which
include overall energy distributions similar to flat-radio-spectrum
quasars (Padovani, Giommi \& Fiore 1997), weak, often broad emission
lines similar to quasars but not seen in FR 1s (Stickel, Fried \& K\"uhr
1993; SFK hereafter and Rector \& Stocke 1998; RS hereafter), and extended
radio structures more luminous than that of FR 1s (Antonucci \& Ulvestad
1985). Also, the $<$$V/V_{max}$$>$ values for the XBLs and RBLs are widely
different (Perlman et al. 1996; RS). Sambruna, Maraschi \& Urry (1996)
explain these differences as due to RBLs and XBLs being selected from
different portions of the overall luminosity function. However, Stickel
(1988a,b), Narayan \& Schneider (1990), and Stocke, Wurtz \& Perlman (1995)
have presented individual examples suggestive of microlensing of
background AGN by stars in foreground galaxies (Ostriker \& Vietri
1986). If a significant fraction of RBLs are not beamed FR 1s, but
rather microlensed quasars, this can also explain the above
differences. However, no convincing individual case of a microlensed BL
Lac has been brought forth as yet.

In this Letter we present statistical evidence for a significant
excess of Mg II absorption systems in the complete RBL sample drawn
from the 1 Jy Survey (K\"uhr et al. 1981). The original sample consisted
of 34 sources and was augmented with 3 new sources by Stickel,
Meisenheimer \& K\"uhr (1994). New redshift information based upon
improved optical spectroscopy by us (RS) and Lawrence et al. (1996)
leave only four 1 Jy BL Lacs without firm or tentative redshifts or lower limits on
redshift due to Mg II absorbers. A total of 10 Mg II absorption systems
have been found so far in the 1 Jy sample.

\section{Observations}

\begin{figure}
\plotone{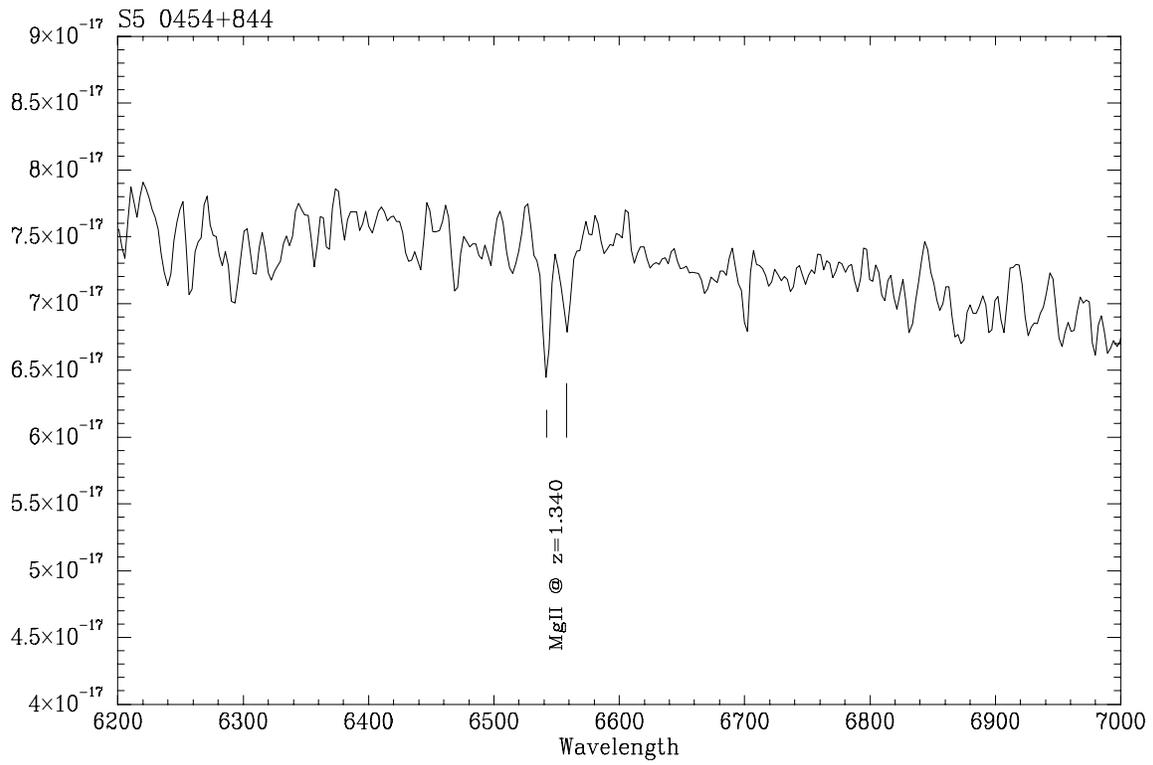}
\caption{A portion of the KPNO 2.1m spectrum of S5 0454+844 showing the
resolution of the absorption line previously detected in a 5m spectrum
by Lawrence et al. (1996) into the Mg II doublet at $z=1.340$.}
\label{fig-1a}
\end{figure}

In order to improve the redshift information in the 1 Jy BL Lac
sample, we obtained new spectroscopy of 1 Jy BL Lacs at the 2.1m
telescope of the Kitt Peak National Observatory and at the Multiple
Mirror Telescope Observatory (MMTO). The 2.1m spectra spanned 4000-8000\AA\ at 8.4\AA\ resolution; the MMTO spectra included 3200-7500\AA\ at
7.4\AA\ resolution. Targets included all 1 Jy objects with no or uncertain
redshift. While the complete spectroscopy will be presented elsewhere
(RS), here we show a portion of a 2.1m spectrum of S5 0454+844,
the MMTO spectrum of PKS 2029+121 and the MMTO spectrum of PKS 0138-097.
The proposed redshift of 0454+844 was based upon the detection of a single
absorption line at $\sim6550$\AA\ by Lawrence et al. (1996), which they
identified as Na ``D" at $z=0.112$. Our spectrum in Figure~\ref{fig-1a} resolves this
line into the Mg II doublet at $z =1.340$. This lower bound on the 
redshift makes this object the most distant now known in the 1 Jy
sample and alters the apparent speed of its jet from subluminal to
superluminal (i.e., $\beta_{app}h > 1.9$c, where $H_0=100h$ km s$^{-1}$
Mpc$^{-1}$ and $q_0=0.5$; Gabuzda \& Cawthorne 1996).  The MMTO spectrum of 2029+121 (Figure~\ref{fig-1b}) reveals emission lines of
C IV, C III] and Mg II at $z=1.215$ with foreground Mg II, Mg I, Fe
II/Mn II 2600,2606\AA\ and Fe II 2382\AA\ absorption at $z=1.117$. Stickel \& K\"uhr (1993) first detected the intervening Mg II absorption system and suggested $z_{em} = 1.223$ based upon only broad Mg II and a tentative detection of C III].   The strength of the newly detected C IV line (rest $W_{\lambda} = 18$\AA) and the Mg II line (rest $W_{\lambda} = 8$\AA) raises some doubts about classifying this object as a BL Lac; although other BL Lacs have been observed with similarly large $W_{\lambda}$ emission lines at some epochs (e.g. 1Jy 1308+326; SFK).  Our new
MMTO spectrum of 0138-097 (Figure~\ref{fig-1c}) confirms the Mg II absorption doublet
previously found by SFK at $z=0.500$ and detects for the first time the
emission line redshift of $z=0.733$ based upon weak Mg II and [O II]. Our
2.1m spectrum of the same object (see RS) detects Mg II, [O II] and
[Ne V] emission as well as Ca II H\&K in absorption, confirming this redshift.

\section{Analysis}

The new redshift information in Lawrence et al. (1996) and RS
added to the SFK data yield the following redshift statistics for the
37 member 1 Jy sample: 23 objects with firm redshifts, 5 objects with tentative redshifts, 4 objects with
only lower bounds on redshift due to the presence of Mg II absorption
and 5 objects with only featureless spectra to date. Of the objects
with firm or tentative redshifts only 16 contribute to the redshift path in which Mg
II absorptions could be found because 12 1 Jy BL Lacs are at $z < 0.4$.
Four other 1 Jy BL Lacs (0820+225, 1144-379, 1519-273 and 2150+173) have not been well-enough observed to support
a limiting equivalent width ($W_{\lambda}$) of less than several
Angstroms and so do not contribute to the observed Mg II pathlength either. Thus, a total of 21 of the 37 1 Jy BL Lac sample
contribute to the Mg II absorption pathlength.

Table~\ref{tbl-1} lists the known Mg II systems in the 1 Jy sample including
10 systems total in 9 different objects with five systems at $W_{\lambda}
> 1$\AA. It is noteworthy that nearly half (4 of 9) of the well-observed
1 Jy BL Lacs without observable emission lines possess Mg II absorbers. Table~\ref{tbl-1} includes the well-known absorption
system in AO 0235+164 at $z=0.524$ (Wolfe \& Wills 1977; Wolfe et al.
1978) and the foreground Mg II absorber due to the lensing galaxy in the
``Smallest Einstein Ring" source B 0218+357 (Browne et al. 1993).

The absorption data in Table~\ref{tbl-1} can be used to compute a Mg II line
density based upon the total Mg II pathlength observed in the entire 1
Jy sample. As mentioned above 21 of the full 37 objects contribute to
the redshift path and most (15), but not all, of these have firm
redshift information. For those few without firm redshifts, we have
estimated ``best redshifts" on the basis of (1) a single emission line
detection (e.g., from SFK or RS) and (2) extended radio source angular
size and luminosity (RS, Murphy, Browne \& Perley 1993 or Antonucci \&
Ulvestad 1985). In addition, for these same sources we set firm upper
and lower bounds to the redshifts using: (1) the Mg II absorbers; (2) the absence of ``host galaxy" in optical images (SFK)
setting $z_{min} = 0.2$; and (3) the absence of ``Ly$\alpha$ forest"
absorption in IUE spectra of three objects, setting $z_{max} = 1.0$ 
(Lanzetta, Turnshek \& Sandoval 1993). In the four objects lacking 
IUE spectra $z_{max}$ was set by the wavelength range of the 
available spectra.

None of these spectra were obtained with the specific purpose of
detecting absorption lines but rather broad, low contrast emission
lines so that moderate resolution spectroscopy (6-18\AA) was employed.
Therefore, only large $W_{\lambda}$ Mg II absorption could have
been detected over the entire Mg II pathlength for a typical BL Lac
spectrum (e.g., those in SFK). For example, the spectrum from SFK in
which the Mg II absorber in 0426-380 was discovered, finds a total
observed $W_{\lambda}$ of $\sim 4$\AA\ yielding a rest-frame $W_{\lambda}$ of 1-1.3\AA\ for the
blue component alone (depending on doublet ratio). Since there are other
possible absorptions in the SFK spectrum of 0426-380 at nearly the same
$W_{\lambda}$, a rest-frame $W_{\lambda}$ limit close to 1\AA\
is indicated. Some spectra have considerably lower $W_{\lambda}$
limits than this one (e.g., those reobserved by Lawrence et al. 1996 or
RS because previous spectra had no detectable emission or absorption
features), but others are quite comparable, so that a 1\AA\ limit is the
best that can be claimed for the sample. Indeed, this limit may
be even overly optimistic for some of the spectra.

With these points in mind, the total pathlength for Mg II
absorption in the 1 Jy sample is 8.1 unit redshifts with a firm lower
bound of 6.8 and maximum value of 11.6. Using the number of Mg II
absorbers with $W_{\lambda} > 1$\AA\ in Table~\ref{tbl-1}, the number density per unit redshift in the 1 Jy sample is: $dN/dz= 0.62 \pm 0.30$, where
the uncertainty is the quadratic sum of the uncertainty in the pathlength and  the sampling statistics.

We compare the above value to the results of the large Mg II
absorption line survey of Steidel \& Sargent (1992) who observed 103
quasars and detected 111 Mg II systems with 36 having $W_{\lambda} >
1$\AA. The best-fit evolution model for the $> 1$\AA\ sample of Steidel
\& Sargent (1992) is: $N(z) = (1+z)^{2.24}$, with a mean redshift for
these absorbers at $z=1.31$. Thus, the $W_{\lambda} > 1$\AA\ absorbers are
mostly detected at higher redshift than the emission line redshift for
all of the 1 Jy BL Lacs. Weighting the number expected at each redshift
by the observed pathlength of the 1 Jy sample we find an expectation
value for the number density of Mg II absorbers in the 1 Jy sample of
$dN/dz=0.14$ systems per unit redshift. Thus, the observed number of high
$W_{\lambda}$ Mg II system is a factor of 4 to 5 times greater than the number expected based upon quasar sightlines; although the uncertainty is large 
due to the sampling statistics.  Because we have found 5 BL Lacs out of 21 with observed Mg II absorption where only 1 was expected, the binomial probability of this occurance is 0.4\% (i.e. 3$\sigma$).  If the maximum observed pathlength of 11.6 is used the probability increases to 1.5\% (2.5$\sigma$).

\section{Discussion}

     The only previous mention of the possibility of an excess of Mg II
absorbers in BL Lac Objects is a brief comment in Weymann, Carswell \&
Smith (1981) that J. Miller (private communication) ``...found Mg II
absorption in what appeared to be an unusually high percentage of such
objects." We confirm Miller's observation with the present statistics
based only on the high $W_{\lambda}$ absorbers in the 1 Jy. But as seen in Table
1, there are a number of lower $W_{\lambda}$ absorbers already found in the 1 Jy
despite the absence of a concerted effort to find them. This excess
could have one of two possible causes:

1. Since four of the ten systems occur in objects lacking emission line
redshifts, these Mg II systems could be intrinsic to the BL Lac.
Aldcroft, Bechtold \& Elvis (1994) have reported an excess of large $W_{\lambda}$
``associated"  Mg II absorbers in radio-loud quasars. If these four
absorbers are associated, this could reduce the excess quoted above,
but not eliminate it entirely since three of the five $W_{\lambda} >
1$\AA\ systems
are clearly foreground to the BL Lac. Therefore, associated absorbers
plus small number statistics could account for the excess. But this
idea  does not explain the correlation between very featureless spectra
and Mg II absorbers.  Nor does the 1 Jy sample contain a single confirmed
associated absorber (i.e., $z_{em} \approx z_{abs}$).

2. For some reason there is a correlation between the presence of a Mg
II absorber along the sightline with a background source whose
characteristics are those of a BL Lac (i.e., a radio-loud AGN with a
featureless or nearly featureless optical spectrum). This possibility
fits under the general hypothesis proposed a few years ago by Ostriker
\& Vietri (1986; OV hereafter) in which the characteristics of a BL Lac
Object are created by the microlensing effect. OV required the
background AGN to be an optically violently variable quasar in order to
understand all the most extreme characteristics of BL Lacs. But the
current statistics do not require all BL Lacs to be microlensed, only
those that show the large $W_{\lambda}$ Mg II absorbers (although the
list in Table~\ref{tbl-1} is certainly not complete since the entire pathlength
foreground to all BL Lacs has not been observed). 

\begin{figure}
\plotone{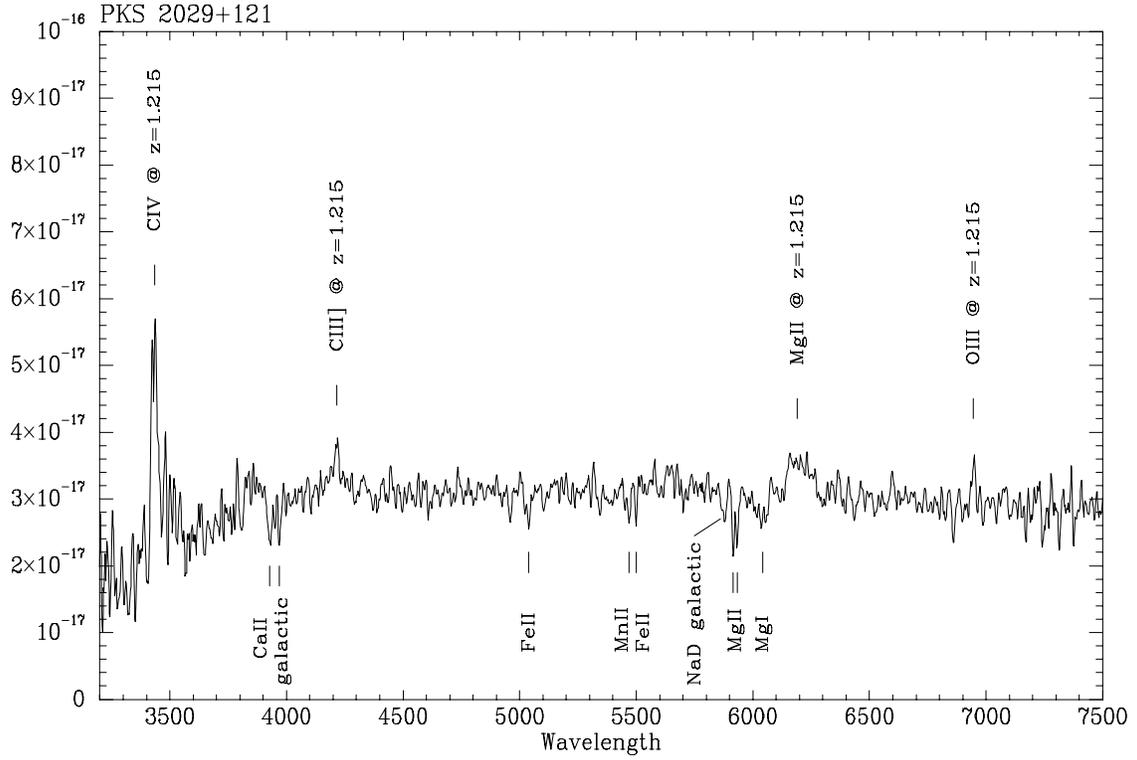}
\caption{The MMTO spectrum of PKS 2029+121 showing the C IV, C III] and Mg
II emission at $z=1.215$, Mg I, Mg II doublet, Mn II 2606\AA, Fe II 2600\AA\
and Fe II 2382\AA\ at $z_{abs}=1.117$ and Galactic Na I ``D" and Ca II H\&K
absorption.}
\label{fig-1b}
\end{figure}

We are aware that there are ample observations (e.g., host galaxies, 
extended radio powers, optical spectra) that strongly support the beamed 
FR 1 hypthesis for low-$z$ BL Lacs, objects which do not 
contribute to the Mg II pathlength.
It is the high-$z$ ($z > 0.5$) BL Lacs that are most discrepant in their properties, many of which have higher radio power levels than FR 1s and weak, quasar-like emission lines in their optical spectra (e.g., see Figure~\ref{fig-1b}); suggesting that many high-$z$ BL Lacs may belong to an
intrinsically different population (or populations).  This issue will be discussed in detail in RS.

Two of the BL Lacs in Table~\ref{tbl-1} have already been sugested as
gravitationally-lensed sources. The case for 0218+357 is clearly made 
in Patnaik et al. (1993) and
Browne et al. (1993). The case of AO 0235+164 is more problematical
since there is no obvious second image and the extended radio structure
is too faint for its morphology to be determined. But the variable
foreground H I absorption is difficult to understand without some
gravitational microlensing (Wolfe et al. 1978, 1982). Abraham et al.
(1993) have used high-resolution ground-based imaging to place
stringent limits on the presence of any second image and use these
limits to argue against the microlensing hypothesis since, by their
estimation, any significant microlensing must be accompanied by
macrolensing that would produce an observable second image (Merrifield
1992). However, Narayan \& Schneider (1990) pointed out that these
constraints are relaxed if the foreground galaxy has low surface mass
density. A test of the microlensing hypothesis for BL Lacs in Table~\ref{tbl-1}
would involve the identification of the Mg II absorber galaxy as a low
surface brightness galaxy like a late-type spiral or irregular. This
type of galaxy is quite different from the types of galaxies found by
Steidel (1995) to be the Mg II absorbers (i.e., $L > 0.1L^{*}$ normal galaxies) in his quasar sample.

\begin{figure}
\plotone{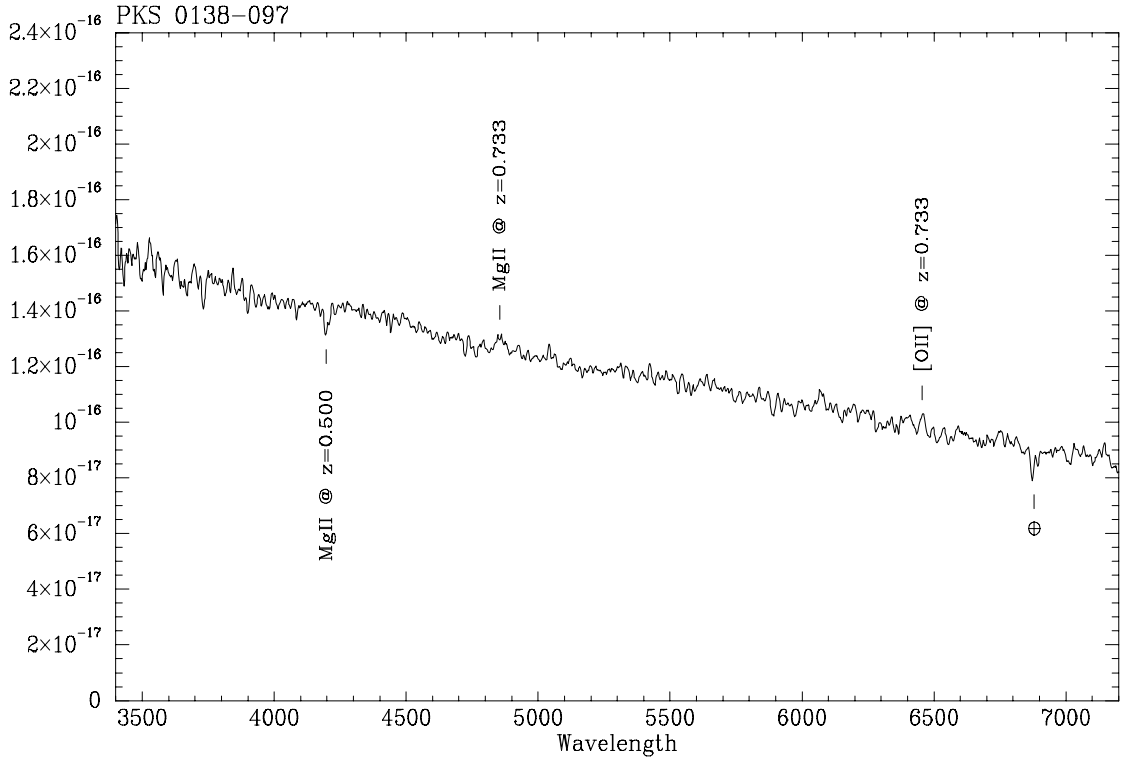}
\caption{The MMTO spectrum of PKS 0138-097 showing the previously known Mg
II absorber at $z=0.500$ (SFK) and the new emission line redshift of $z=0.733$ from Mg II and possible [O II]. A 2.1m spectrum (RS) confirms the Mg II and [O II] emission, while detecting [Ne V] in addition at the same redshift.}
\label{fig-1c}
\end{figure}

In this proposed scenario, BL Lac Objects with foreground Mg II
absorbers are microlensed by stars associated with the absorbing gas.
The background AGN is a radio-loud quasar with normal emission line
properties but the action of the foreground stellar screen is to
preferentially amplify the continuum emission region relative to the
broad-line region, creating an optical spectrum like that in Figure~\ref{fig-1c}.
Microlensing also explains the correlation between the presence of the
Mg II absorber and the very featureless BL Lac spectra. 

If this hypothesis is correct, the largest unanswered question is
why AGN with ``featureless" spectra are always radio loud (Stocke et al.
1990). Also, if this hypothesis is correct, conclusions concerning BL
Lacs which have relied upon complete 1 Jy sample properties must be
reevaluated (e.g., Urry \& Padovani 1995).

Finally, we note the great similarity between the BL Lac spectrum of
PKS 0138-097 shown in Figure~\ref{fig-1c} and the optical spectrum of the
$\gamma$-ray burst source GRB 970508 recently obtained by Metzger et al. 
(1997).  This similarity and the abundance of BL Lacs amongst known,
bright $\gamma$-ray sources suggests that the transient gravitational
microlensing of AGN could account for $\gamma$-ray bursters.
If $\gamma$-ray bursts are lensed AGN, this would greatly reduce the
intrinsic energy requirements of the $\gamma$-ray bursts through both
relativistic beaming and gravitational lensing.

\acknowledgments
Research on BL Lac Objects at the University of Colorado is supported by NASA grant NAGW-2675.


\clearpage
\begin{deluxetable}{lrrcl}
\tablecaption{Mg II Absorption Systems In 1 Jy BL Lac Objects \label{tbl-1}}
\tablewidth{0pt}
\tablehead{\colhead{} & \colhead{} & \colhead{} & \colhead{$W_{\lambda}$ (2796\AA)} &
\colhead{} \\
\colhead{Object} & \colhead{$z_{em}$} & \colhead{$z_{abs}$} &
\colhead{(Rest Frame)} & \colhead{Reference}}
\startdata
0118-272   &  \nodata   &    0.559    &     0.8    &  Falomo (1991) \\

0138-097   &  0.733     &    0.500    &     0.3    &  This Paper \\

0218+357   &  0.94:     &    0.685    &     1.8    &  Browne et al. (1993) \\

0235+164   &  0.94      &    0.852    &     0.4    &  Wolfe \& Wills (1977) \\
           &            &    0.524    &     2.4    &  \\
 
0426-380   &  \nodata   &    1.030    &    1-1.3   &  SFK \\

0454+844   &  \nodata   &    1.340    &     0.4    &  This Paper \\

0735+178   &  \nodata   &    0.424    &     1.1    &  Carswell et al. (1974) \\

1308+326   &  0.996     &    0.879    &     0.4    &  Briggs \& Wolfe (1983) \\

2029+121   &  1.215     &    1.117    &     1.5    &  This Paper \\
\enddata

\end{deluxetable}

\clearpage

\clearpage

\clearpage

\end{document}